\documentclass[showpacs, showkeys, aps]{revtex4-1}

\usepackage{amssymb,amsmath,xcolor,graphicx,xspace,colortbl,textcomp, rotating} 
\usepackage{boxedminipage}  
\usepackage{fixltx2e}  
\usepackage{ragged2e}  
\usepackage{tabulary}  
\usepackage{xcolor}  
\graphicspath{{LY14964A-PRA-Kajita_graphics/}{LY14964A-PRA-Kajita_tcache/}{LY14964A-PRA-Kajita_gcache/}}
\DeclareGraphicsExtensions{.pdf,.eps,.ps,.png,.jpg,.jpeg}

\begin{document}
\title{Accuracy estimation of the \textsuperscript {16}O\textsubscript {2}\textsuperscript {+} transition frequencies targeting the search for the variation in $m_{p}$/$m_{e}$ 
}
\author{Masatoshi Kajita}
\email[E-mail me at: ]{kajita@nict.go.jp}
\homepage[Visit: ]{http://www.nict.go.jp/en/index.html}
\affiliation{National Institute of Information and Communications Technology,  Koganei, Tokyo 184-8795, Japan
}
\date{
\today
}
\begin{abstract}\begin{description}
\item [Abstract]In this paper, we estimate the Stark and Zeeman shifts in the transition frequencies of the \textsuperscript {16}O\textsubscript {2}\textsuperscript {+ }molecular ion,  as a step for the search for the variation in the proton-to-electron mass ratio $\mu $. The $X^{2}\Pi $ $v =21 -a^{4}\Pi $ $v =0$ or the $X^{2}\Pi v =21 -a^{4}\Pi $ $v =1$  transition frequencies (THz region) of the \textsuperscript {16}O\textsubscript {2}\textsuperscript {+} molecular ion have particularly high sensitivity to the variation in $\mu $. Note also that the Stark shift in the \textsuperscript {16}O\textsubscript {2}\textsuperscript {+ }transition frequencies is expected to be much smaller than that for heteronuclear diatomic molecules. However, the actual systematic uncertainties for the \textsuperscript {16}O\textsubscript {2}\textsuperscript {+} transition frequencies have never been estimated. We estimated the Stark and Zeeman shifts in the different \textsuperscript {16}O\textsubscript {2}\textsuperscript {+} transition frequencies. When the molecular ions in a string crystal formed in a linear trap (trap electric field \textless{} 0.1 V/cm, and Stark shift \textless{} 10\textsuperscript {-20}) are used, the $X$\textsuperscript {2}$\Pi _{1/2}(v ,J) =(0 ,1/2) -(v\rq ,1/2)(v\rq$$ \geq 1)$ transition frequencies are most advantageous for the search for the variation in $\mu \left (\Delta \mu /\mu  <10^{ -17}\right )$ because the Zeeman shift is easily suppressed to lower than 10\textsuperscript {-18} and the electric quadrupole shift is zero. On the other hand, the $X^{2}\Pi _{1/2}\left (v ,J\right ) =\left (21 ,1/2\right ) -a^{4}\Pi _{1/2}\left (v ,J\right ) =\left (0 ,1/2\right )$ transition frequency has another merit that the positive Stark shift induced by the trap electric field can be canceled by the quadratic Doppler shift. Therefore, the measurement using molecular ions in a Coulomb crystal broadened in the radial direction is also possible, when the Zeeman shift is effectively eliminated.
\end{description}
\end{abstract}
\maketitle

\section{Introduction
}
   The standard model of physics is based on the assumption that the fundamental constants are perfectly constant throughout time and space, but Dirac mentioned the possibility of their variation in 1937 \cite{Dirac}. If there are variations in some fundamental constants, energy structure of atoms and molecules will change. The variation of fundamental constants can be evaluated by measuring the variation in the ratio of two transition frequencies with different sensitivities. The variation in the fine structure constant $\alpha $ has become one of the hottest subjects for researchers since the measurement uncertainties of some atomic transition frequencies were reduced to lower than 10\textsuperscript {-15} \cite{Bize}\cite{Huntemann1}\cite{Ushijima}\cite{Nicholson}\cite{Chou}. Currently, the upper limit of the variation in $\alpha $ is estimated to be less than 10\textsuperscript {-17}/yr  \cite{Rosenband}\cite{Godun}\cite{Huntemann-alpha}.

 The variation in the proton-to-electron mass ratio $\mu  =m_{p}/m_{e}$ has also been investigated because the ratio of the variations in $\alpha $ and $\mu $ provides useful information for grand unification theories \cite{Calmet1}. Comparing the Cs hyperfine transition frequency with the \textsuperscript {171}Yb\textsuperscript {+}  transition frequency, Godun et al. and Huntemann et al. found that $\Delta \mu /\mu $ cannot be larger than $1$$0$\textsuperscript {-16}/yr \cite{Godun}\cite{Huntemann-alpha}.  Note that the variation in this frequency ratio reflects not only the variation in $\mu $ but also that in $\alpha $. To evaluate the model-independent variation in $\mu $, it is preferable to measure the variation in the ratio of the molecular vibrational-rotational transition frequencies to the \textsuperscript {1}S\textsubscript {0}-\textsuperscript {3}P\textsubscript {0} transition frequencies of a  \textsuperscript {87}Sr atom or an \textsuperscript {27}Al\textsuperscript {+ }ion, which have very low sensitivity to $\mu $ and $\alpha $ \cite{Angstmann} and measurement uncertainty less than 10\textsuperscript {-17} \cite{Ushijima}\cite{Nicholson}\cite{Chou}. However, molecular transitions have never been measured with an uncertainty lower than 10\textsuperscript {-14}, although a stability of $6 \times 10^{ -15}$ was obtained with an I\textsubscript {2} -stabilized diode laser \cite{Philippe}.  

The precise measurement of molecular transitions is difficult because the complicated quantum energy structure makes laser cooling and localization in a selected quantum state difficult. Diatomic molecules including an atom without nuclear spin have relatively simple energy structures. The vibrational transition frequencies of X\textsuperscript {6}Li molecules in an optical lattice  \cite{Kajita-YbLi}\cite{Kajita-YbLi2}\cite{Kajita-CaLi} and XH\textsuperscript {+} molecular ions in a linear trap \cite{Kajita-CaH+}\cite{Kajita-XH+} are considered to have been measured with uncertainty lower than 10\textsuperscript {-16}, where X is the even isotope of a group II atom (\textsuperscript {40}Ca, \textsuperscript {88}Sr etc.). However, the production of X\textsuperscript {6}Li molecules with kinetic energy lower than 10 $\mu $K has never been attained. XH\textsuperscript {+} molecular ions have been produced, and the overtone-vibrational transition of a \textsuperscript {40}CaH\textsuperscript {+} molecular ion has been observed \cite{Khanyile}. The main problem in the case of using XH\textsuperscript {+} molecular ions is that they have a large permanent electric dipole moment and their internal states are rearranged through the interaction with black-body radiation (BBR). A cryogenic environment is required to mitigate this effect.

A cryogenic chamber is not required for measurement using homonuclear diatomic molecular ions because there is no electric-dipole (E1) transition between different vibrational-rotational states in the electronic ground state. The Stark shift is much smaller than that for heteronuclear diatomic molecules because it is induced only by the coupling with electronic excited states. The N\textsubscript {2}\textsuperscript {+}  $(I =0)$ $\left (v ,N\right ) =\left (0 ,0\right ) \rightarrow \left (v^{ \prime } ,0\right )$ $\left (v^{ \prime } \geq 1\right )$ transitions are expected to be measured with uncertainty lower than 10\textsuperscript {-17}  (\textit{$I$}: nuclear spin, $v$: vibrational state, $N$: rotational state) because the Zeeman and quadrupole shifts are zero \cite{Kajita-N2+a}\cite{Kajita-N2+b}. The \textsuperscript {14}N\textsubscript {2}\textsuperscript {+} $(v ,N) =(0 ,0) -(1 ,2)$ transition has actually been observed \cite{Germann-a}. N\textsubscript {2}\textsuperscript {+} molecular ions are produced in a selected vibrational-rotational state by resonance-enhanced multiphoton ionization (REMPI) \cite{REMPI-b}. To prepare the \textsuperscript {14}N\textsubscript {2}\textsuperscript {+} molecular ion with $I =0$, high resolution REMPI is required, while $I$ is always $0$ for the \textsuperscript {15}N\textsubscript {2}\textsuperscript {+} molecular ion (natural abundance of 1 ppm) with $N =0$.

Also the \textsuperscript {16}O\textsubscript {2}\textsuperscript {+} molecular ion is an attractive homonuclear diatomic molecular ion because the \textsuperscript {16}O nuclear spin is zero. Hanneke et al. proposed measuring the \textsuperscript {16}O\textsubscript {2}\textsuperscript {+} $X^{2}\Pi $ $v =21 -a^{4}\Pi $ $v =0$ or $X^{2}\Pi v =22 -a^{4}\Pi $ $v =1$ transition frequencies for the reason given below \cite{Hanneke}. The energy of the high vibrational state $E_{Xv}/h \approx 1000$ THz is approximately proportional to $\mu ^{ -0.5}$, while the energy of the $a^{4}\Pi v =0 ,1$ state $E_{a}$ has almost no dependence on $\mu $. The variation in $E_{\delta }/h =\left (E_{a} -E_{Xv}\right )/h$ induced by the variation in $\mu $ is given by $\Delta E_{\delta } = -0.5E_{Xv}\left (\Delta \mu /\mu \right ) = -PE_{\delta }\left (\Delta \mu /\mu \right )$, where $P =0.5 \times E_{Xv}/E_{\delta }$. The $X^{2}\Pi v =21$ and $a^{4}\Pi v =0$ ($X^{2}\Pi v =22$ and $a^{4}\Pi $) states are accidently quasi-degenerated and $E_{\delta }/h$ s very sensitive to the variation in $\mu $ because $P >100$. However, the possibility of searching for the variation in $\mu $ should be discussed not only interest of the sensitivity but also interest of the attainable frequency accuracy. The systematic frequency uncertainty for different shifts have never been estimated for the \textsuperscript {16}O\textsubscript {2}\textsuperscript {+} transition frequencies.   In this paper, we discuss the Stark and Zeeman shifts in the different \textsuperscript {16}O\textsubscript {2}\textsuperscript {+} vibrational-rotational transition frequencies, including the  $X^{2}\Pi $ $v =21 -a^{4}\Pi $ $v =0$ transition frequency. Considering both the sensitivity to the variation in $\mu $ and the attainable frequency measurement accuracy, the $X^{2}\Pi _{1/2}$ $\left (v ,J\right ) =\left (0 ,1/2\right ) -\left (v^{ \prime } ,1/2\right )\left (v \geq 1\right )$ transition frequencies are most advantageous for the search for the variation in $\mu $ when molecular ions in a string crystal formed in a linear trap (trap electric field \textless{} 0.1 V) are used. Here, $J$ is the total angular momentum given by the electron spin, electron orbital angular momentum, and molecular rotation.  

Measurement of the $X^{2}\Pi _{1/2}\left (v ,J\right ) =\left (0 ,1/2\right ) -\left (v^{ \prime } ,1/2\right )\left (v \geq 1\right )\text{}$ transition frequencies can be performed with simpler experimental apparatus than that used for the measurement of the transition frequencies proposed by Hanneke et al. \cite{Hanneke}. The production of \textsuperscript {16}O\textsubscript {2}\textsuperscript {+} molecular ions in the desired vibrational rotational state with $v \leq 1$ has already been realized by REMPI \cite{Dochain}, but it is more difficult to prepare molecular ions in the highly excited state. The $v =0 -v^{ \prime }$ transition is probed with a laser in the infrared or optical region, whose frequency stabilization is much easier than that for a THz wave to detect the transitions between accidentally degenerated states. Note also that the $X^{2}\Pi  -a^{4}\Pi $ transition rate is much lower than that for the vibrational transitions in the $X^{2}\Pi $ state \cite{Hanneke}. The quantum state of the molecular ion after the irradiation of the probe laser is monitored by a quantum logical detection system \cite{Quantum logic}.

\section{Sensitivity to the variation in fundamental constants
}

When a fundamental constant changes by $X \rightarrow X +\Delta X$, the atomic or molecular transition frequency changes by $f \rightarrow f +\Delta f$. The parameter $\lambda _{X}$ used to show the sensitivity of $f$ to the variation in $X$ is defined by\begin{equation}\lambda _{X} =\frac{\left (\Delta f/f\right )}{\left (\Delta X/X\right )} =\frac{X}{f}\frac{df}{dX} . \label{lamda}
\end{equation}Values of $\lambda _{X}$ are estimated from the change in $f$ obtained by the ab-initio calculation upon changing the value of $X$ slightly. When $f$ is given as a simple function of $X$, $\lambda _{X}$ can also be obtained from the simple formula for $\left (df/dX\right )$. By measuring the ratio of the two transition frequencies $f_{1 ,2}$ with different values of $\lambda _{X}\left (\lambda _{X1 ,2}\right )$, the variation in $X$ is obtained by evaluating

 \begin{equation}\left (\Delta X/X\right ) =\frac{1}{\left (\lambda _{X1} -\lambda _{X2}\right )}\frac{\Delta \left (f_{1}/f_{2}\right )}{\left (f_{1}/f_{2}\right )} .
\end{equation}

 Detection of the variation in $X$ is possible when $\Delta f$  is larger than the frequency measurement uncertainty $\delta f$. Therefore, the minimum detectable value of $\left (\Delta X/X\right )$ is estimated as $\left \vert \delta f/\left (\lambda _{X}f\right )\right \vert $, although Ref. \cite{Hanneke} discusses only $\lambda _{X}$.

 For a model-independent search of the pure variation in $\mu $, the ratios of molecular vibrational-rotational transition frequencies to a reference frequency with small $\lambda _{\mu }$ and $\lambda _{\alpha }$ should be measured. The \textsuperscript {87}Sr \textsuperscript {1}S\textsubscript {0}-\textsuperscript {3}P\textsubscript {0} transition frequency (429.2 THz, 698 nm) is one of the best references since $\lambda _{\mu } <10^{ -4}$, $\lambda _{\alpha } =0.06$ \cite{Angstmann}, and the frequency uncertainty is on the order of 10\textsuperscript {-18} \cite{Ushijima}\cite{Nicholson}.  

Table 1 shows the \textsuperscript {16}O\textsubscript {2}\textsuperscript {+} $X^{2}\Pi _{1/2}$ $v =0 \rightarrow v^{ \prime }$ ($v^{ \prime } =1 ,4 ,8$) transition frequencies $f$ and the corresponding values of  $\lambda _{\mu }$. These transition frequencies are given by $f =v^{ \prime }f_{v} -v^{ \prime }\left (v^{ \prime } +1\right )xf_{v}$ with a difference of less than 0.01\% from the experimental result \cite{Song}, where $f_{v}$ (=57.1 THz) and $xf_{v}$ (=0.487 THz ) are the harmonic and the unharmonic vibrational potential terms, respectively. Since $f_{v} \propto \mu ^{ -0.5}$ and $xf_{v} \propto \mu ^{ -1}$ , $\lambda _{\mu }$ is approximately given by $\left [ -0.5v^{ \prime }f_{v} +v^{ \prime }\left (v^{ \prime } +1\right )xf_{v}\right ]/f$. The effect of the variation in the rotational constant on the estimation of $\lambda _{\mu }$ is negligible for the vibrational transition frequency.  The  $v =0 \rightarrow 4$ transition is convenient for cooperative measurement between two laboratories, because the probe laser light (1369 nm) can be transferred to a distant location via an optical fiber and can be compared with a \textsuperscript {87}Sr lattice clock laser (698 nm) after frequency doubling. The $v =0 \rightarrow 8$ transition is convenient for direct comparison with a \textsuperscript {87}Sr lattice clock laser with a frequency difference of 7.3 THz. The values of $f$ and $\lambda _{\mu }$ are also shown for the  \textsuperscript {16}O\textsubscript {2}\textsuperscript {+} $X^{2}\Pi _{1/2}$ $v =21 \rightarrow a^{4}\Pi _{1/2}v =0 $transition, which was proposed by Hanneke et al. \cite{Hanneke} because of the large value of $\lambda _{\mu }$.

\begin{tabular}[c]{|l|l|l|}\hline
Transition
&
$f$(THz) \cite{Song}
&
$\lambda _{\mu }$
\\
\hline
$X^{2}\Pi _{1/2}$ $v =0 \rightarrow 1$
& 56.5
& -0.49
\\
\hline
$X^{2}\Pi _{1/2}$ $v =0 \rightarrow 4$
& 219.0
& -0.48
\\
\hline
$X^{2}\Pi _{1/2}$ $v =0 \rightarrow 8$
& 421.9
& -0.46
\\
\hline
$X^{2}\Pi _{1/2}$ $v =21 \rightarrow a^{4}\Pi _{1/2}v =0$
& 2.7
& 140
\\
\hline
\end{tabular}\\
Table 1: \textsuperscript {16}O\textsubscript {2}\textsuperscript {+} transition frequencies $f$ and sensitivity parameters for the proton-to-electron mass ratio $\lambda _{\mu }$.
\\

\section{Estimation of the Stark shift
}
For homonuclear diatomic molecules, the Stark shift is induced only by the coupling with electronically excited states and its dependence on the rotational state is very small.  Using the values of the transition frequencies and Einstein coefficients listed in Ref. \cite{N2Table}, the dc quadratic Stark energy shift in each vibrational state in the $X^{2}\Pi $  $v =0 ,1 ,4 ,8 ,$ $J =1/2$ state ($a^{4}\Pi $ $v =0 ,$ $J =1/2$ state) was obtained by considering the coupling with the $A^{2}\Pi $ $v =0 -21$ ($b^{4}\Sigma $ $v =0 -7$) states. Table 2 lists the dc quadratic Stark coefficients  $\delta f_{S}/\left [\lambda _{\mu }fE^{2}\right ]$ for each transition frequency ($E$: electric field). For the molecular ions in a string crystal formed in a linear trap, the trap electric field is less than 0.1 V and  $\vert \delta f_{S}/\left [\lambda _{\mu }f\right ]\vert  <10^{ -20}$ is attained.

A Stark $\delta f_{BBR}$ is also induced by BBR, which is approximately proportional to $T^{4}$, where $T$ is the ambient temperature. Table 2 lists $\delta f_{BBR}/\left [\lambda _{\mu }f\left (T\text{(K)}/300\right )^{4}\right ]$. Stabilizing $T$ by replacing it with $T \pm \Delta T$, the uncertainty of $\delta f_{BBR}/\left [\lambda _{\mu }f\right ]$ is reduced by a factor of $4\Delta T/T$.

The Stark shift is also induced by a probe laser and is proportional to the laser intensity $I_{p}$. Values of  $\delta f_{P}/\left [\lambda _{\mu }fI_{p}\right ]$, where $\delta f_{P}$ is the Stark shift induced by the probe laser, are also shown in Table 2. Observing the one-photon quadrupole (E2) transition, the probe laser intensity is expected to be less than 10  mW/cm\textsuperscript {2} and $\delta f_{P}/\left [\lambda _{\mu }f\right ] <10^{ -18}$.  This shift can be further suppressed by the hyper-Ramsey method \cite{Hyper-Ramsey1}.

Comparing $\delta f_{S}/\left [\lambda _{\mu }fE^{2}\right ]$, $\delta f_{BBR}/\left [\lambda _{\mu }f\left (T\left (\text{K}\right )/300\right )^{4}\right ]$, and $\delta f_{P}/\left [\lambda _{\mu }fI_{p}\right ]$, there is no significant difference between the different \textsuperscript {16}O\textsubscript {2}\textsuperscript {+} transition frequencies listed in Table 2. However, note that the positive Stark shift induced by the trap electric field can be eliminated by cancellation with the quadratic Doppler shift (see Section 5) by applying a suitable rf-trap electric field frequency \cite{Dube}. For the $X^{2}\Pi _{1/2}v =21 \rightarrow a^{4}\Pi _{1/2}v =0$ transition frequency,  $\delta f_{S}$ is positive and it can be eliminated by applying rf- the trap electric field with the frequency of 11 MHz. For the $X^{2}\Pi _{1/2}v =0 \rightarrow v^{ \prime }$ transition frequencies, $\delta f_{S}$ is negative ( $\delta f_{S}/\left [\lambda _{\mu }fE^{2}\right ]$ shown in Table 2 is positive with negative values of $\lambda _{\mu }$) and the cancellation with the quadratic Doppler shift is not possible.

\begin{tabular}[c]{|l|l|l|l|}\hline
\multicolumn{1}{|c|}{Transition
} & $\delta f_{S}/\left [\lambda _{\mu }fE^{2}\right ]$(/(V/cm)\textsuperscript {2})
& $\delta f_{BBR}/\left [\lambda _{\mu }f\left (T\left (\text{K}\right )/300\right )^{4}\right ]$
& $\delta f_{P}/\left [\lambda _{\mu }fI_{p}\right ]$
\par (/(W/cm\textsuperscript {2})
\\
\hline
$X^{2}\Pi _{1/2}$ $v =0 \rightarrow 1$ & $1.1 \times 10^{ -19}$ & $7.0 \times 10^{ -18}$ & $4.3 \times 10^{ -17}$ \\
\hline
$X^{2}\Pi _{1/2}$ $v =0 \rightarrow 4$ & $6.2 \times 10^{ -20}$ & $4.0 \times 10^{ -18}$ & $2.4 \times 10^{ -17}$ \\
\hline
$X^{2}\Pi _{1/2}v =0 \rightarrow 8$  & $6.8 \times 10^{ -20}$ & $4.3 \times 10^{ -18}$ & $3.4 \times 10^{ -17}$ \\
\hline
$X^{2}\Pi _{1/2}$ $v =21 \rightarrow a^{4}\Pi _{1/2}$ $v =0$ & $8.9 \times 10^{ -20}$ & $5.6 \times 10^{ -18}$ & $3.3 \times 10^{ -17}$ \\
\hline
\end{tabular}\\ Table 2: Stark coefficients induced by a dc electric field, $\delta f_{S}/E^{2}$, blackbody radiation $\delta f_{BBR}/(T/300)^{4}$ , and a probe laser $\delta f_{P}/I_{p}$ listed as ratios to $\lambda _{\mu }f$ to show the utility of searching for the variation in the proton-to-electron mass ratio $\mu $.\\

\section{Estimation of Zeeman shift
}
    The Zeeman shift in the \textsuperscript {16}O\textsubscript {2}\textsuperscript {+} molecular ion in the $X^{2}\Pi _{\Omega }$ $\left (v ,J\right )$ state is estimated by a considerably different method from that in molecules in the $\Sigma $ state because the electron orbital angular momentum and electron spin are defined with the component parallel to the molecular axis with the quantum numbers $\Lambda $ and $\Sigma $, respectively. Here, $\Omega  =\Lambda  +\Sigma $ and there is an energy gap $A_{v}$ between different $\Omega $ states at each vibrational state. The molecular rotation is not defined by an independent quantum number (with the $\Sigma $ state, defined by $N$), and the rotational energy is given by the total angular momentum $J$ as $B_{v}$ $\left [J\left (J +1\right ) -\Omega ^{2}\right ]$, where $B_{v}$ is the rotational constant in each vibrational state. With the non-relativistic approximation, the linear Zeeman energy shift $E_{Z}$ in the $X^{2}\Pi _{\Omega }$ $\left (v ,J\right )$ state is given by \cite{Townes}

\begin{equation}E_{Z} =\genfrac{(}{)}{}{}{\mu _{B}}{h}\frac{M}{J\left (J +1\right )}\left [\Omega \left (g_{L}\Lambda  +g_{S}\Sigma \right ) +g_{R(v ,\Omega  ,J)}\left (J\left (J +1\right ) -\Omega ^{2}\right )\right ]B \label{Zeeman11}
\end{equation}
where $B$ is the magnetic field, $M$ is the component parallel to the magnetic field, and $\mu _{B}$ is the Bohr magneton ($\mu _{B}/h =1.3996$ MHz/G). The g-factors of electron orbital angular momentum, electron spin, and molecular rotation are denoted as $g_{L}\left ( =1\right )$, $g_{S}\left ( =2.002\right )$, and $g_{R\left (v ,\Omega  ,J\right )}( =3.06 \times 10^{ -5}$ with $v =0 ,$ $\Omega  =J =1/2)$, respectively. While $g_{L}$ and $g_{S}$ have no dependence on the vibrational state, $g_{R(v ,\Omega  ,J)}$ (estimated by the method shown in Ref. \cite{Karr}) has a dependence on $v$, $\Omega $, and $J$. $E_{Z}$ does not change with the $\Lambda  \rightarrow  -\Lambda  ,\Sigma  \rightarrow  -\Sigma  ,\Omega  \rightarrow  -\Omega $ transforms. The \textsuperscript {16}O\textsubscript {2}\textsuperscript {+} nuclear spin is zero, therefore, there is no hyperfine structure. 

The Zeeman shift in the transition frequency $\delta f_{Z}$ is given by the difference between $E_{Z}$ in the upper and lower states. Considering $\Omega $ as a ``good quantum number``,
$\left (\Lambda  ,\Sigma \right )$ is (1,1/2) in the $X^{2}\Pi _{3/2}$ state and (1,-1/2) in the $X^{2}\Pi _{1/2}$ state. $E_{Z}/B$ in the $X^{2}\Pi _{3/2}$ state is on the order of $ \pm $1 MHz/G, while it is less than $ \pm $1 kHz/G in the $X^{2}\Pi _{1/2}$ state. Therefore, $\delta f_{Z}/B$ in the $X^{2}\Pi _{1/2} \rightarrow X^{2}\Pi _{3/2}$ transition frequency is on the order of $ \pm $1 MHz/G and is not suitable for precise measurement. Therefore we only consider the $\Delta \Omega  =0$ transitions. 

Table 3 shows the linear Zeeman coefficients in the  $X^{2}\Pi _{1/2}$ $v =0 \rightarrow 1$ vibrational-rotational transition frequencies. The Zeeman shift in the  $\Delta J =\Delta M =0$ transition frequency is much less than that in the $\Delta J =2$ transition frequency because of the cancellation of the shifts in the upper and lower states. To examine the Zeeman shift in the $\Delta \Omega  =\Delta J =\Delta M =0$ transitions, the energy structure should be considered in more detail. There are off-diagonal matrix elements of the Hamiltonian $ -B_{v}\sqrt{J\left (J +1\right ) -3/4}$ between the $X^{2}\Pi _{3/2}$ $\left (v ,J\right )$ and the $X^{2}\Pi _{1/2}$ $\left (v ,J\right )$ states \cite{Hanneke}, and they induce a mixture of both states ($\Omega  -$mixture). For the $\Delta J =\Delta M =0$ $J \geq 3/2$ transition, the $\Omega  -$mixture leads the significant dependence of $g_{R(v ,\Omega  ,J)}$ on $v$, $\Omega $, and $J$ \cite{Sauer}. The dependence of $g_{R(v ,\Omega  ,J)}$ on $v$ is mainly given by the dependence of $A_{v}$ and $B_{v}$ on $v$ ($A_{0} =6.00$ THz, $B_{0} =50.4$ GHz and $A_{1} =5.98$ THz, $B_{1} =49.8$ GHz \cite{Song}). Because of the $X^{2}\Pi _{1/2} -X^{2}\Pi _{3/2}$ coupling, there is a quadratic Zeeman shift with a coefficient smaller than $ \pm $0.1 Hz/G\textsuperscript {2} for the $J \geq 3/2$ states. The Zeeman shift in the $X^{2}\Pi _{3/2}\left (v ,J ,M\right ) =\left (0 ,J^{ \prime } ,M^{ \prime }\right ) \rightarrow \left (v^{ \prime } ,J^{ \prime } ,M^{ \prime }\right )$ transition frequency is $\left ( -1\right ) \times $ (Zeeman shift in the $X^{2}\Pi _{1/2}$ $\left (0 ,J^{ \prime } ,M^{ \prime }\right ) \rightarrow \left (v^{ \prime } ,J^{ \prime } ,M^{ \prime }\right )$ transition frequency).

The $\Omega  -$mixture effect does not exist for the $X^{2}\Pi _{1/2}J =1/2$ state, and $|(g_{R(1,1/2,1/2)}-g_{R(0,1/2,1/2)})/g_{R(0,1/2,1/2)}| (\approx 0.01)$ is much smaller than $\vert \left (g_{R(1 ,\Omega  ,J)} -g_{R(0 ,\Omega  ,J)}\right )/g_{R\left (0 ,\Omega  ,J\right )}\vert $ with $J \geq 3/2$. The relativistic effect leads a dependence of the spin-orbit interaction on $v$, but this effect is negligible small (correction ratio $ <10^{ -10}$) because the nuclear vibrational motion velocity is less than 2000 m/s. The linear Zeeman coefficient in the $X^{2}\Pi _{1/2}$ $\left (v ,J ,M\right ) =\left (0 ,1/2 , \pm 1/2\right ) \rightarrow \left (v^{ \prime } ,1/2 , \pm 1/2\right )$ transition frequency is much smaller than that in the $\Delta J =\Delta M =0 ,J \geq 3/2$ transition frequency. The Zeeman shift in the  $\left (J ,M\right ) =\left (1/2 , \pm 1/2\right ) \rightarrow \left (1/2 , \pm 1/2\right )$ transition is strictly linear with the magnetic field, and the Zeeman shift is perfectly eliminated by averaging the $M = \pm 1/2 \rightarrow  \pm 1/2$ transition frequencies. Considering that $g_{R\left (v^{ \prime } ,1/2 ,1/2\right )} -g_{R(0 ,1/2 ,1/2)}$ is approximately proportional to $v^{ \prime }$ \cite{Karr}, the change in  $\delta f_{Z}/\left [\lambda _{\mu }fB\right ]$ is less than 10\% for the $X^{2}\Pi _{1/2}$ $\left (v ,J ,M\right ) =\left (0 ,1/2 , \pm 1/2\right ) \rightarrow \left (v^{ \prime } ,1/2 , \pm 1/2\right )$ transition frequencies with $v^{ \prime } =1 -8$.

\begin{tabular}[c]{|l|l|l|}\hline
\multicolumn{1}{|c|}{
$X^{2}\Pi _{1/2}\left (v ,J ,M\right )$
} &
$\delta f_{Z}/B$ (Hz/G)
&
$\delta f_{Z}/\left [\lambda _{\mu }fB\right ]$(/G)
\\
\hline
$\left (0 ,1/2 , \pm 1/2\right ) \rightarrow \left (1 ,1/2 , \pm 1/2\right )$
& $ \mp 0.14$
&
$ \pm 5.1 \times 10^{ -15}$
\\
\hline
$\left (0 ,3/2 , \pm 3/2\right ) \rightarrow \left (1 ,3/2 , \pm 3/2\right )$
&
$ \mp 4.3$
&
$ \pm 1.6 \times 10^{ -13}$
\\
\hline
$\left (0 ,5/2 , \pm 5/2\right ) \rightarrow \left (1 ,5/2 , \pm 5/2\right )$
&
$ \mp 14$
&
$ \pm 5.2 \times 10^{ -13}$
\\
\hline
$\left (0 ,1/2 , \pm 1/2\right ) \rightarrow \left (1 ,5/2 , \pm 5/2\right )$
&
$ \pm 1100$
&
$ \mp 4.1 \times 10^{ -11}$
\\
\hline
\end{tabular}\\
Table 3: Linear Zeeman coefficients in the $X^{2}\Pi _{1/2}$ $v =0 \rightarrow 1$ vibrational-rotational transition frequencies $\delta f_{Z}/B$ and their ratio to $\lambda _{\mu }f$.
\\

In the $a^{4}\Pi $ state, the linear Zeeman coefficient is smallest in the $a^{4}\Pi _{1/2}$ state, and it is on the order of $ \pm 1$ MHz/G in other $a^{4}\Pi _{\Omega }$ states. The couplings between different $\Omega $ states are much larger than those between the $X^{2}\Pi _{3/2}$ and $X^{2}\Pi _{1/2}$ states because of the smaller $A_{v}$ ($A_{0} = -1.4$ THz \cite{Hanneke}). This effect is also significant when $J =1/2$ because of the coupling between the $a^{4}\Pi _{1/2}$ and  $a^{4}\Pi _{ -1/2}$ states. For the $X^{2}\Pi _{1/2}$ $\left (v =21 ,J =1/2 ,M = \pm 1/2\right ) \rightarrow a^{4}\Pi _{1/2}$ $\left (v =0 ,J =1/2 ,M = \pm 1/2\right )$ transition frequency, $\delta f_{Z}/B = \pm 1.4$ kHz/G and $\delta f_{Z}/\left [\lambda _{\mu }fB\right ] = \pm 3.8 \times 10^{ -12}$ /G.  There is also a quadratic Zeeman shift with a coefficient smaller than $ \pm $1 Hz/G\textsuperscript {2}. Therefore,  the $X^{2}\Pi _{1/2}$ $\left (v ,J ,M\right ) =\left (0 ,1/2 , \pm 1/2\right ) \rightarrow \left (v^{ \prime } ,1/2 , \pm 1/2\right )$ transition frequencies are most advantageous in the search for the variation in $\mu $ since they suppress the Zeeman shift to less than $10^{ -18}$.

\section{Other frequency uncertainties}

For the molecular ions in a linear trap, there is a significant electric field gradient and the electric quadrupole shift can be a serious problem in precise measurement. This shift is proportional to $3M^{2} -J\left (J +1\right )$, and it is zero for the $J =1/2$ state. Measurement of the $J =1/2 \rightarrow 1/2$ transition frequency is also advantageous for this reason. For other cases, the electric quadrupole shift should be eliminated by averaging the transition frequencies with different $M$.

The quadratic Doppler shift $\delta f_{QD}$ is proportional to the kinetic energy $K$ and $\delta f_{QD}/fK = -1/\left (m_{i}c^{2}\right ) = -4.4 \times 10^{ -18}$/mK for all transition frequencies ($m_{i}$: mass of molecular ion). Then $\delta f_{QD}/f\lambda _{\mu }K =8.8 \times 10^{ -18}$/mK for the pure vibrational frequencies and $\delta f_{QD}/f\lambda _{\mu }K = -3.1 \times 10^{ -20}$/mK for the $X^{2}\Pi _{1/2}v =21 \rightarrow a^{4}\Pi _{1/2}v =0$ transition frequency. A kinetic energy lower than 0.1 mK can be obtained by sideband Raman cooling and $\vert \delta f_{QD}/f\lambda _{\mu }\vert  <10^{ -18}$ can be attained. 

The gravity redshift is given by $\delta f_{G}/fH =g_{a}/c^{2} =10^{ -18}$/cm, where $H$ is the altitude and $g_{a}$ is the acceleration due to gravity. For the pure vibrational transition frequency, $\delta f_{G}/f\lambda _{\mu }H = -2 \times 10^{ -18}$/cm, and for the $X^{2}\Pi _{1/2}v =21 \rightarrow a^{4}\Pi _{1/2}v =0$ transition frequency,  $\delta f_{G}/f\lambda _{\mu }H =7.1 \times 10^{ -21}$/cm.

\section{Comparison with other homonuclear diatomic molecular ions\textsuperscript {}
}
  Reference \cite{Kajita-N2+a} showed that the N\textsubscript {2}\textsuperscript {+} $\left (I =0\right )X^{2}\Sigma $ $\left (v ,N ,J ,M\right ) =\left (0 ,0 ,1/2 , \pm 1/2\right ) \rightarrow \left (v^{ \prime } ,0 ,1/2 , \pm 1/2\right )$ transition frequency can be measured with uncertainty lower than $10^{ -17}$. In this section, we compare this transition frequency with the \textsuperscript {16}O\textsubscript {2}\textsuperscript {+} $X^{2}\Pi _{1/2}$ $\left (v ,J ,M\right ) =\left (0 ,1/2 , \pm 1/2\right ) \rightarrow \left (v^{ \prime } ,1/2 , \pm 1/2\right )$ transition frequency.

The main difference is that the rotational energy is zero in the N\textsubscript {2}\textsuperscript {+} $N =0$ state, while it is nonzero with the \textsuperscript {16}O\textsubscript {2}\textsuperscript {+} $J =1/2$ state. Because of the zero rotational energy, the N\textsubscript {2}\textsuperscript {+} transition frequency can be observed without a Zeeman shift, and uncertainty lower than 10\textsuperscript {-17} can be obtained with a single transition. For the \textsuperscript {16}O\textsubscript {2}\textsuperscript {+} transition frequency, the rotational energy is nonzero, which makes the Zeeman shift nonzero. Uncertainty lower than 10\textsuperscript {-17} can be obtained by averaging the $M = \pm 1/2 \rightarrow  \pm 1/2$ transition frequencies. A two-photon transition is required for the N\textsubscript {2}\textsuperscript {+} transition, while the one-photon E2 transition is possible for the \textsuperscript {16}O\textsubscript {2}\textsuperscript {+} transition.

The preparation of molecular ions with $I =0$ is necessary for measurement using N\textsubscript {2}\textsuperscript {+} molecular ion \cite{Kajita-N2+b}. For this purpose, high-resolution REMPI is required with \textsuperscript {14}N\textsubscript {2}\textsuperscript {+}. Another method is to use \textsuperscript {15}N\textsubscript {2}\textsuperscript {+}, whose natural abundance is on the order of 1 ppm. The \textsuperscript {16}O nuclear spin is zero; therefore, the preparation of \textsuperscript {16}O\textsubscript {2}\textsuperscript {+} molecular ions in $X^{2}\Pi _{1/2}$ $\left (v ,J\right ) =\left (0 ,1/2\right )$ by REMPI is possible using a simple laser system and natural O\textsubscript {2} gas \cite{Dochain}. Therefore, measurement using \textsuperscript {16}O\textsubscript {2}\textsuperscript {+ }molecular ion is possible with a simpler apparatus than that with N\textsubscript {2}\textsuperscript {+}.

Measurement of the vibrational transition frequencies of H\textsubscript {2}\textsuperscript {+}, D\textsubscript {2}\textsuperscript {+}, and HD\textsuperscript {+} molecular ions is useful to obtain the absolute values of $\mu $ and other parameters because they can be calculated by solving the Schroedinger and Dirac equations analytically \cite{Karr2}\cite{Schiller}. The energy structures of H\textsubscript {2}\textsuperscript {+} and D\textsubscript {2}\textsuperscript {+} molecular ions are the same as those of the \textsuperscript {15}N\textsubscript {2}\textsuperscript {+} and \textsuperscript {14}N\textsubscript {2}\textsuperscript {+} molecular ions, respectively. However, the measurement of the vibrational transition frequencies with uncertainty lower than 10\textsuperscript {-17} is difficult, mainly because of the significant quadratic Doppler shift (for a kinetic energy of 1 mK, $7.0 \times 10^{ -17}$ for H\textsubscript {2}\textsuperscript {+} and $3.5 \times 10^{ -17}$ for D\textsubscript {2}\textsuperscript {+}). For the measurement of the HD\textsuperscript {+} transition frequency, the complicated hyperfine structure makes it difficult to localize the molecular ion in a selected quantum state.

\section{
Conclusion}
This paper discussed the possibility of searching for the variation in the proton-to-electron mass ratio $\mu $ via precise measurement of the transition frequencies of \textsuperscript {16}O\textsubscript {2}\textsuperscript {+} molecular ions, considering both the sensitivity parameter for the variation in $\mu $ ($\lambda _{\mu }$) and the attainable accuracy ($\delta f/f$). Using molecular ions in a string crystal formed in a linear trap (trap electric field \textless{} 0.1 V/cm, and Stark shift \textless{} 10\textsuperscript {-20}),  the \textsuperscript {16}O\textsubscript {2}\textsuperscript {+} $X^{2}\Pi _{1/2}$ $\left (v ,J ,M\right ) =\left (0 ,1/2 , \pm 1/2\right ) \rightarrow \left (v^{ \prime } ,1/2 , \pm 1/2\right )$ transition frequencies are most advantageous in searching for the variation in $\mu $ with $\Delta \mu /\mu  <10^{ -17}$, because the Zeeman shift is strictly linear with small coefficients ($\delta f_{Z}/\lambda _{\mu }fB = \pm 5.1 \times 10^{ -15}$ /G) and can be eliminated perfectly by averaging the $M = \pm 1/2 \rightarrow  \pm 1/2$ transition frequencies. Note also that the electric quadrupole shift is zero.

The $X^{2}\Pi _{1/2}$ $\left (v ,J ,M\right ) =\left (21 ,1/2 , \pm 1/2\right ) \rightarrow a^{4}\Pi _{1/2}$ $\left (v ,J ,M\right ) =\left (0 ,1/2 , \pm 1/2\right )$ transition has a large value of $\lambda _{\mu }$, as proposed in Ref. \cite{Hanneke}. However, it is not an advantageous transition for obtaining a low $\left [\delta f/\lambda _{\mu }f\right ]$, because the linear Zeeman shift is much larger than that for the $X^{2}\Pi _{1/2}\left (v ,J\right ) =\left (0 ,1/2\right ) \rightarrow \left (v^{ \prime } ,1/2\right )$ transition frequencies ($\delta f_{Z}/\lambda _{\mu }fB = \pm .83 \times 10^{ -12}$ /G) and also the quadratic Zeeman shift exists. However, this transition has the merit that the Stark shift induced by the trap electric field is positive and can cancel with the quadratic Doppler shift when the rf-trap electric field frequency is 11 MHz \cite{Dube}. Therefore, measurement using molecular ions in a string crystal is not required. Also, the electric quadrupole shift is zero for this transition frequency.

Measurement of the $X^{2}\Pi $ $v =0 -v^{ \prime }$ transition frequencies is much easier than that for the $X^{2}\Pi v =21 \rightarrow a^{4}\Pi $ $v =0$ or $X^{2}\Pi v =22 \rightarrow a^{4}\Pi v =1$ transition frequencies because (1) the preparation of \textsuperscript {16}O\textsubscript {2}\textsuperscript {+} molecular ions in the $X^{2}\Pi v =0 ,J =1/2$ state is much easier than that in the highly excited state \cite{Dochain}, (2) frequency stabilization of the probe laser in the infrared or optical region is much easier than that for a THz wave to induce the transition between accidentally degenerated states,  and (3) the vibrational transition rate is much higher than that of the $X^{2}\Pi  -a^{4}\Pi $ transition.
\begin{acknowledgments}This research was supported by a Grant-in-Aid for Scientific Research (B) (Grant No. JP25287100), a Grant-in-Aid for Scientific Research (C) (Grant No. JP16K05500), and a Grant-in-Aid for Exploratory Research (Grant No. JP15K13545) from the Japan Society for the Promotion of Science (JSPS). I greatly appreciate the fruitful discussion with C. Shi and P. O. Schmidt (PTB).
\end{acknowledgments}

\subparagraph{
}

\subparagraph{
}

\subparagraph{
}

\end{document}